\documentclass[twocolumn, pra]{revtex4-1} 
\usepackage{amssymb} 
\usepackage{amsmath}
\usepackage{epsfig}
\usepackage{graphics, graphicx}
\usepackage{bbold}
\usepackage{psfrag}
\usepackage{mathcomp}
\usepackage{subfigure}
\usepackage{verbatim}
\usepackage{color}
\usepackage[colorlinks,citecolor=blue]{hyperref}
\def\cp#1{\mathbf{#1}}

\begin{document}

\title{Visualizing the dispersions of Fermi polaron and molecule via spin-orbit coupling}
\author{Tingting Shi}
\affiliation{Beijing National Laboratory for Condensed Matter Physics, Institute of Physics, Chinese Academy of Sciences, Beijing, 100190, China}
\author{Xiaoling Cui}
\email{xlcui@iphy.ac.cn}
\affiliation{Beijing National Laboratory for Condensed Matter Physics, Institute of Physics, Chinese Academy of Sciences, Beijing, 100190, China}
\date{\today}

\begin{abstract}
We propose to measure the dispersions of Fermi polaron and molecule by engineering spin-orbit coupling (SOC) on the impurity, which induces spin flip with finite momentum transfer. The polaron dispersion can be probed at small SOC momentum from the linear response of impurity spin. For molecule, we show that it can be prepared through an adiabatic steady-state evolution when setting  SOC momentum as the Fermi momentum of majority bath. By gradually reducing SOC strength to zero, the steady state smoothly evolves to a molecular state with directional symmetry breaking. The corresponding dispersion can then be probed experimentally through the pair momentum distribution of molecules at finite density. Our scheme reveals a fundamental momentum difference between Fermi polaron and molecule, thereby offering a clear physical picture for their first-order transition in single-impurity system.
\end{abstract}
\maketitle

\section{Introduction}

As a typical quasi-particle, Fermi polaron describes the collective motion of an impurity dressed by interactions with surrounding majority fermions. Owing to the high controllability of ultracold atoms, Fermi polaron has been successfully realized in highly polarized fermion mixtures\cite{Zwierlein,Salomon,Grimm,Kohl,Grimm2016,Roati,Zwierlein2,Sagi,Grimm2021,Grimm2024}, where both attractive and repulsive  branches have been explored. Key quasiparticle properties, including energy, residue, and effective mass, have been measured via
 direct\cite{Zwierlein,Kohl,Zwierlein2} or inverse\cite{Grimm,Grimm2021,Grimm2024,Roati} radio-frequency (rf) spectroscopy, conducted in the linear response regime at short time and small rf coupling. Going beyond this regime, a recent experiment has developed a steady-state spectroscopy for Fermi polarons driven by a strong rf field\cite{Navon}, where the original polaron picture is severely destroyed after a long-time relaxation.

Despite these developments, an important phenomenon of Fermi polarons, i.e., polaron-molecule transition\cite{Prokofev, Leyronas, Punk, Enss, Bruun, Castin, Parish,Parish2, Houcke, MC_2d_1, MC_2d_2}, remains unobserved experimentally. This transition occurs in the attractive branch of both three-dimensional (3D) and two-dimensional (2D) polarons, whose ground state transits from a polaron (an impurity dressed by many particle-hole excitations) to a molecule  (a bound state between the impurity and a single majority fermion outside the Fermi sea) as  the impurity-fermion attraction increases. In literature, such a first-order transition  has been predicted by various theoretical approaches including  diagrammatic Monte Carlo\cite{Prokofev, Houcke, MC_2d_1, MC_2d_2}, functional renormalization group\cite{Enss}, variational\cite{Punk, Castin, Parish,Parish2} and T-matrix\cite{Leyronas,Bruun} methods\cite{footnote}.  In these studies,  polaron and molecule states are treated separately, either by adopting different variational ansatz\cite{Punk, Castin, Parish,Parish2} or  by extracting their energies from distinct quantities in one-body or two-body channel\cite{Prokofev,Leyronas,Bruun,Houcke, MC_2d_1, MC_2d_2,Enss}. 
Later, a unified variational ansatz for both states was proposed by generalizing Chevy's ansatz \cite{Chevy} to arbitrary momentum\cite{Cui,Cui2}. The resulting energy dispersion exhibits two minima: one at zero momentum (${\cp Q} = 0$) and another at the Fermi momentum of the majority fermions ($|{\cp Q}| = k_F$), corresponding to the polaron and molecule states, respectively. In this framework, the polaron-molecule transition can be interpreted as an energy competition between different momentum states, which acquires a first-order character. 

However, in realistic finite-temperature systems with finite impurity density and finite temperature, such a first-order transition is smoothed into a continuous crossover due to the coexistence of polaron and molecule near their transition\cite{Sagi,Cui, Cui2, Parish3}. Crossing the transition point, the impurities smoothly evolve from polarons at ${\cp Q}\sim 0$ to molecules  at finite ${\cp Q}$, with amplitude $|{\cp Q}|\sim k_F$ but direction free.  During this process, the total momentum of impurities remains zero, which therefore does not reflect the important momentum difference between polaron and molecule in basic single-impurity system.  Here we note that a transparent way to distinguish polaron and molecule states and to visualize their first-order transition is through a direct measurement of  impurity dispersion, particularly the double-minimum structure near the transition point. We note that previous theoretical studies have examined the impurity dispersion\cite{HH2022} and molecule preparation/spectrum\cite{Schmidt} from Raman spectroscopy. In comparison, here we aim to extract both the polaron and molecule dispersions, with particular emphasis on their intrinsic momentum difference.

\begin{figure}[t]
\includegraphics[width=8.5cm]{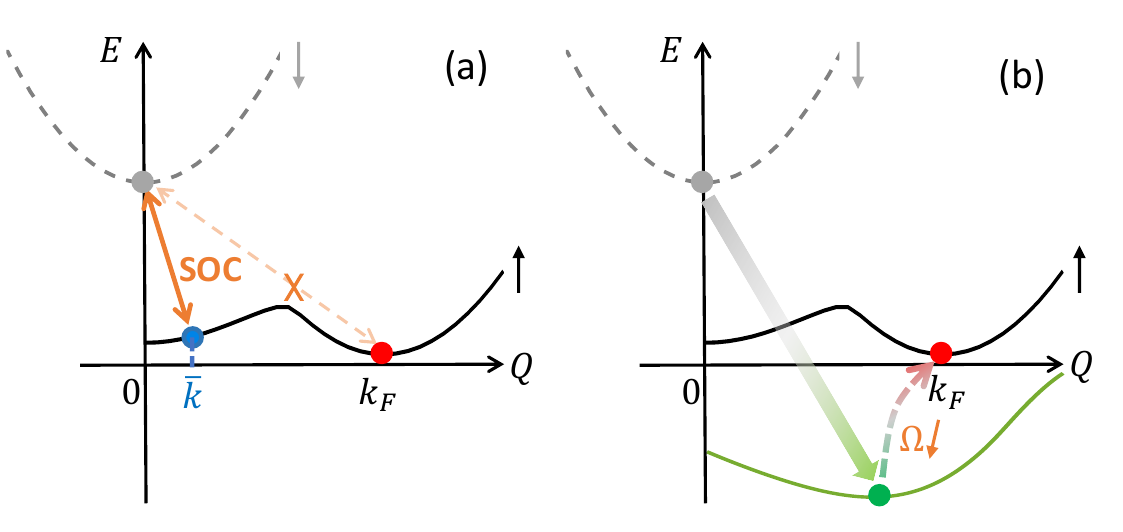}
\caption{(Color Online). Illustration of using spin-orbit coupling (SOC) to detect the dispersions of Fermi polaron (a) and molecule (b). For a specific example we take the SOC momentum along x direction, $\bar{\cp k}=\bar{k}{\cp e}_x$, and only show energy dispersion along x. In (a),  SOC with small momentum  $\bar{k}$ ($\ll k_F$) is applied to couple the initially non-interacting impurity (gray point) with Fermi polaron at finite momentum $Q(=\bar{k})$ (blue point). 
Polaron energy $E_Q$ can be detected through the linear response of impurity spin at short time. This scheme, however, fails to detect molecule at $Q\sim k_F$ because of vanishingly small effective coupling induced by SOC. An alternative way to detect molecule  is illustrated in (b), where $\bar{k}$ is set to be $k_F$ and the impurity is initially relaxed to the ground (steady) state at large SOC strength $\Omega$ (green point). As gradually reducing $\Omega$, the steady state evolves adiabatically and eventually at $\Omega\rightarrow 0$ recovers molecule state (red point). The dispersion near $Q\sim k_F$ can then be probed through  the pair momentum distribution of finite-density molecules.}  \label{fig_schematic}
\end{figure}

In this work, we propose to measure the impurity dispersion by engineering a spin-orbit coupling (SOC)\cite{soc_review} on the impurity, which induces spin flip with a controlled momentum transfer  ($\bar{\cp k}$). As illustrated in Fig.\ref{fig_schematic}(a), the polaron dispersion at small  ${\cp Q}$ can be probed with SOC momentum $\bar{\cp k}={\cp Q}$, by detecting   short-time dynamics of  impurity spin in the linear response regime. In the limit $\bar{\cp k}=0$, this scheme reduces to standard inverse rf spectroscopy\cite{Grimm,Grimm2021,Grimm2024,Roati}. However, this linear response approach fails to detect molecule at $|{\cp Q}|\sim k_F$, due to its vanishing quasi-particle residue\cite{Cui,Cui2} and thus infinitesimal impurity-molecule coupling produced by SOC. Instead, we propose to prepare the molecular state via an adiabatic steady-state evolution under SOC with $|\bar{\cp k}|=k_F$, see Fig.\ref{fig_schematic}(b). Starting from a unique ground (steady) state at large SOC strength, the system is adiabatically evolved to zero SOC. This process continuously connects the initial state to a molecular state with specific momentum ${\cp Q}=\bar{\cp k}$ that breaks the rotational symmetry. In a realistic gas, the molecule dispersion can then be measured from the pair momentum distribution of finite-density molecules.
We remark that our scheme is particularly useful for probing the double-well structure of impurity dispersion when polaron and molecule are both locally stable. The distinct $\bar{\cp k}$ used for the two states clearly reflect their momentum difference, thereby offering a clear physical picture for their first-order transition in single-impurity system.

The rest of this paper is organized as follows. Section~\ref{model} introduces the model and formalism for our single-impurity system with and without spin-orbit coupling. Section~\ref{polaron} addresses the detection of polarons at small momentum within the linear-response regime, and Section~\ref{molecule} focuses on the preparation of molecules via adiabatic state evolution. We conclude with a summary of our work in Section~\ref{summary}.

\section{Model and Formulism} \label{model}

We start with the following Hamiltonian  for a spin-orbit coupled impurity imbedded in a 3D Fermi gas ($\hbar=1$):
\begin{eqnarray}
H&=&\sum_{\cp k} \left[  \epsilon_{\cp k}  c^{\dag}_{{\cp k} \uparrow}c_{{\cp k} \uparrow}+ (\epsilon_{\cp k}+\delta) c^{\dag}_{{\cp k} \downarrow}c_{{\cp k} \downarrow} +\Omega (c^{\dag}_{{\cp k}+\bar{\cp k};\uparrow}c_{{\cp k} \downarrow}+h.c.) \right] \nonumber\\
&&+\sum_{\cp k}\epsilon^f_{\cp k} f^{\dag}_{\cp k}f_{\cp k} +\frac{g}{V}\sum_{{\cp Q},{\cp k},{\cp k'}} c^{\dag}_{{\cp k}\uparrow}f^{\dag}_{{\cp Q}-{\cp k}}f_{{\cp Q}-{\cp k'}}c_{{\cp k'}\uparrow}. \label{H}
\end{eqnarray}
Here $c^{\dag}_{{\cp k} \sigma}$ creates a spin-$\sigma$($\uparrow,\downarrow$) impurity with kinetic energy  $\epsilon_{\cp k}={\cp k}^2/(2m)$ and detuning $\delta$; the impurity is dressed by spin-orbit coupling (SOC) with strength $\Omega$ and transferred momentum $\bar{\cp k}$ --- both are practically tunable by the intensity and propagation direction of two-photon Raman lasers\cite{soc_review};   $f^{\dag}_{\cp k}$ creates a majority fermion with energy $\epsilon^f_{\cp k}={\cp k}^2/(2m_f)$; $g$ is the bare interaction between the fermion and spin-$\uparrow$ impurity, which can be renormalized via $1/g=\mu/(2\pi a_s)-1/V\sum_{\cp k}(2\mu/{\cp k}^2)$ with s-wave scattering length $a_s$ and reduced mass $\mu=mm_f/(m+m_f)$. In this work we consider equal mass with $m_f=m$ and $\mu=m/2$. Extension to unequal-mass case will be discussed at the end. 

In the absence of SOC  ($\Omega=0$),  (\ref{H}) describes the usual Fermi polaron system, for which we can write down a varitional ansatz\cite{Chevy, Punk, Castin, Cui}:
\begin{equation}
|P_{\cp Q}\rangle=\left(\phi^{\uparrow}_0 c^{\dag}_{{\cp Q}\uparrow}+\sum'_{{\cp k},{\cp q}}\phi^{\uparrow}_{{\cp k}{\cp q}} c^{\dag}_{{\cp Q}+{\cp q}-{\cp k},\uparrow}f^{\dag}_{\cp k}f_{\cp q}\right)  |{\rm FS}\rangle_{N}, \label{P_Q}
\end{equation}
where $|{\rm FS}\rangle_N$ is the Fermi-sea state of majority atoms with number $N$, giving the Fermi  momentum $k_F$ and Fermi energy $E_F=k_F^2/(2m)$;  $\sum'$ refers to the momentum summation with constraint $|\cp k|>k_F$ and  $|\cp q|\leqslant k_F$; ${\cp Q}$ is the total momentum in the reference frame of   $|{\rm FS}\rangle_{N}$; the bare coefficient is $\phi^{\uparrow}_0=\sqrt{Z_{\cp Q}}$ with $Z_{\cp Q}$ the quasi-particle residue. Based on the general ansatz (\ref{P_Q}), it has been shown that the polaron-molecule transition originates from an energy competition between ${\cp Q}=0$ and $|{\cp Q}|=k_F$ states\cite{Cui, Cui2}, with the latter well incorporating the bare molecule  ansatz with zero center-of-mass momentum\cite{Punk}.

When turn on SOC ($\Omega\neq 0$), a variational ansatz for the eigen-state of (\ref{H}) can be written as:
\begin{eqnarray}
&&|\bar{P}_{\cp Q}\rangle=\left(\bar{\phi}^{\uparrow}_0 c^{\dag}_{{\cp Q}\uparrow}+\sum'_{{\cp k},{\cp q}}\bar{\phi}^{\uparrow}_{{\cp k}{\cp q}} c^{\dag}_{{\cp Q}+{\cp q}-{\cp k},\uparrow}f^{\dag}_{\cp k}f_{\cp q}\right. \nonumber\\
 &&\ \ \ \left.+\bar{\phi}^{\downarrow}_0 c^{\dag}_{{\cp Q}-\bar{\cp k};\downarrow}+\sum'_{{\cp k},{\cp q}}\bar{\phi}^{\downarrow}_{{\cp k}{\cp q}} c^{\dag}_{{\cp Q}-\bar{\cp k}+{\cp q}-{\cp k},\downarrow}f^{\dag}_{\cp k}f_{\cp q} \right) |{\rm FS}\rangle_{N}.  \label{wf_p}
\end{eqnarray}
Compared to (\ref{P_Q}), (\ref{wf_p}) contains additional terms in its second line due to inter-spin coupling via  SOC.   For simplicity we have truncated  terms in (\ref{P_Q},\ref{wf_p}) up to the lowest order with one particle-hole excitations. Its validity will be discussed later.

By imposing the Schr{\"o}dinger equation $H|\bar{P}_{\cp Q}\rangle=\bar{E}_{\cp Q}|\bar{P}_{\cp Q}\rangle$, one can extract the eigen-energy $\bar{E}_{\cp Q}$ from a self-consistent equation and further obtain all variational coefficients in (\ref{wf_p}). A similar problem has been solved previously\cite{Yin}, focusing on equilibrium properties of spin-orbit coupled polarons. Instead, here we aim at using SOC to facilitate the exploration of polaron/molecule dispersions in original impurity system (i.e., in the absence of SOC). In particular, we are most interested in the interaction regime where both polaron and molecule are locally stable, which corresponds to $1/k_Fa_s$ near the transition point ($=1.27$ up to the lowest-order truncation in (\ref{P_Q})\cite{Cui,Leyronas,Punk}). 

\section{Polaron detection at small ${\cp Q}$}  \label{polaron}

First, we point out that  SOC can be directly used to explore the polaron dispersion at small momentum ${\cp Q}=\bar{\cp k}$. In fact, in the limit of $\bar{\cp k}=0$, SOC reduces to Rabi (or rf) coupling, which has been widely adopted in ultracold experiments to detect zero-momentum polaron (${\cp Q}=0$)\cite{Zwierlein,Kohl,Zwierlein2, Grimm,Grimm2021,Grimm2024,Roati}. Take the inverse rf spectroscopy for example, starting from an initially non-interacting impurity ($c^{\dag}_{{\cp 0} \downarrow}$), a rf field flips its spin and couples it with zero-momentum polaron  $|P_{0}\rangle$, whose energy and residue can then be detected through the dynamical response of impurity spin\cite{Grimm,Grimm2021,Grimm2024,Roati}. Similarly, SOC with a finite $\bar{\cp k}$ can transfer the initial impurity $c^{\dag}_{{\cp 0} \downarrow}$ to $c^{\dag}_{\bar{\cp k} \uparrow}$ and  couple it with  $|P_{\bar{\cp k}}\rangle$, see  Fig.\ref{fig_schematic}(a). Consequently, the energy and residue of finite-momentum polaron $|P_{\cp Q}\rangle$ can be directly probed through the dynamical response of impurity spin by applying  SOC with a given $\bar{\cp k}={\cp Q}$. In the following we will take a specific $\bar{\cp k}$ along $x$ direction, i.e., 
\begin{equation}
\bar{\cp k}=\bar{k}{\cp e}_x, 
\end{equation}
which also selects  the polaron momentum ${\cp Q}$ along this direction as ground state. Thus we can simplify $\bar{\cp k}$ and ${\cp Q}$ by their projections  $\bar{k}$ and $Q$ respectively. 

To quantify the response of impurity spin under SOC, we have computed the time-evolution of impurity magnetization $M\equiv (N_{\uparrow}-N_{\downarrow})/(N_{\uparrow}+N_{\downarrow})$ under the time-dependent wavefunction:
\begin{equation}
|\Psi(t)\rangle=\sum_j c_j e^{-i\bar{E}^{(j)}_{\bar{k}} t} |\bar{P}^{(j)}_{\bar{k}} \rangle, \label{psi_t}
\end{equation}
where $|\bar{P}^{(j)}\rangle$ is the $j$-th eigen-state (Eq.\ref{wf_p}) with eigen-energy $\bar{E}^{(j)}$;    $c_j=\langle \bar{P}^{(j)}_{\bar{k}}|\Psi(t=0)\rangle$, with the initial state $|\Psi(t=0)\rangle=c^{\dag}_{0 \downarrow}|{\rm FS}\rangle_N$. 
In our calculation we have summed up as many eigen-states as possible in attractive and repulsive branches. On the other hand, the problem can be greatly simplified if the detuning $\delta$ is only close to one eigen-state  of original system but far from others. Here we consider $\delta$ close to the attractive polaron  energy ($E_{\bar{k}}$). In this case,  $H$ can be expanded as  a  $2\times 2$ matrix in the bases of $c^{\dag}_{0 \downarrow}|{\rm FS}\rangle_N$ and  $|{P_{\bar{k}}}\rangle$:
\begin{equation}
H=\left(\begin{array}{cc} \delta & \Omega_{\rm eff} \\ \Omega_{\rm eff} & E_{\bar{k}}\end{array}\right), \label{H_22}
\end{equation}
with off-diagonal term given by an effective coupling 
\begin{equation}
\Omega_{\rm eff}=\sqrt{Z_{\bar{k}}}\Omega. \label{Omega}
\end{equation}
The time evolution of $M$ can then be straightforwardly obtained. In the linear response regime where $\Omega_{\rm eff} t\ll 2\pi$, $M$ displays the fastest growth at $\delta=E_{\bar{k}}$, which can be directly used to probe the polaron dispersion at finite momentum $Q=\bar{k}$. 

\begin{figure}[t]
\includegraphics[width=8.5cm]{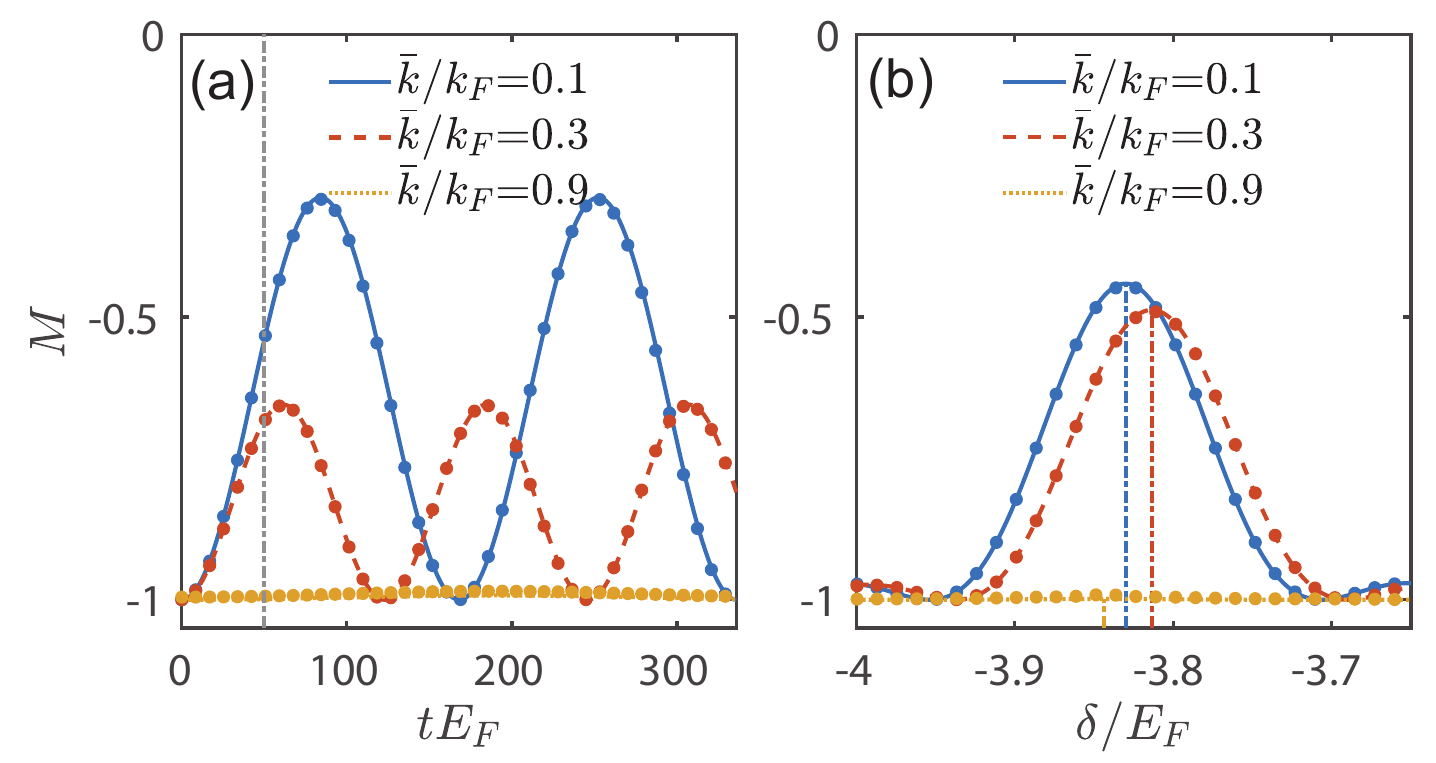}
\caption{(Color Online). Linear response of impurity spin under spin-orbit coupling (SOC) at $\Omega/E_F=0.03$ and $1/(k_Fa_s)=1.3$. (a) Time evolution of impurity magnetization $M(t)$ at a fixed detuning $\delta=-3.86E_F$ and different SOC momenta $\bar{k}/k_F=0.1,\ 0.3,\ 0.9$.  (b)  $M$ at a given short time $t_0=50/E_F$ as a function of $\delta$. $M(t_0)$ shows a maximum at $\delta=E_{\bar{k}}$, as located by vertical lines for various $\bar{k}$. In (a,b), circles and lines  are respectively from numerical evaluations using (\ref{psi_t}) and simplified two-level model  (\ref{H_22}). }  \label{fig_Ep}
\end{figure}

Taking $1/(k_Fa_s)=1.3$ and a small $\Omega=0.03E_F$, in Fig.\ref{fig_Ep}(a,b) we show the dynamical response of $M$ at different $\bar{k}$. Given $\delta$ very close to the attractive polaron energy ($E_{\bar{k}}$),  the numerical results based on (\ref{psi_t}) match well with the predictions from (\ref{H_22}).  In particular, for small $\bar{k}\ (\ll k_F)$, $M$ displays a visible response even at short time $t\ll2\pi/\Omega_{\rm eff}$.  At a given short time $t=t_0$, the response is indeed most pronounced at $\delta=E_{\bar{k}}$, as located by vertical lines in Fig.\ref{fig_Ep}(b). This confirms the feasibility of using SOC to probe polaron dispersion at small $Q=\bar{k}$ in the linear response regime. 

However, this scheme fails for detecting molecule state at large $Q=\bar{k}\sim k_F$, as shown by exceedingly weak response of $M$ in Fig.\ref{fig_Ep}(a,b) for $\bar{k}=0.9k_F$. Physically, this can be attributed to  the infinitesimal effective coupling $\Omega_{\rm eff}$ in (\ref{Omega}) due to vanishingly small residue $Z_{Q}$ at large $Q\sim k_F$\cite{Cui, Cui2}.  As a result, the molecule state can hardly respond to SOC in the linear response regime. 

\section{Molecule preparation at  $|{\cp Q}|\sim k_F$}  \label{molecule}

To detect molecule, we suggest to  first prepare it through an adiabatic steady-state evolution under SOC with $\bar{k}=k_F$. In fact, a recent experiment has realized steady-state spectroscopy in Rabi-coupled Fermi polarons\cite{Navon}. 
Similarly, here one can first prepare the ground (steady) state with a strong SOC field (large $\Omega$), which resides at a single minimum with momentum $k_{\min}$. As illustrated in Fig.\ref{fig_schematic}(b), by gradually reducing $\Omega$, the steady state continuously evolves and ultimately ends in $k_{\min}=k_F$ as $\Omega\rightarrow 0$, yielding the molecule state.

\begin{figure}[t]
\includegraphics[width=8.5cm]{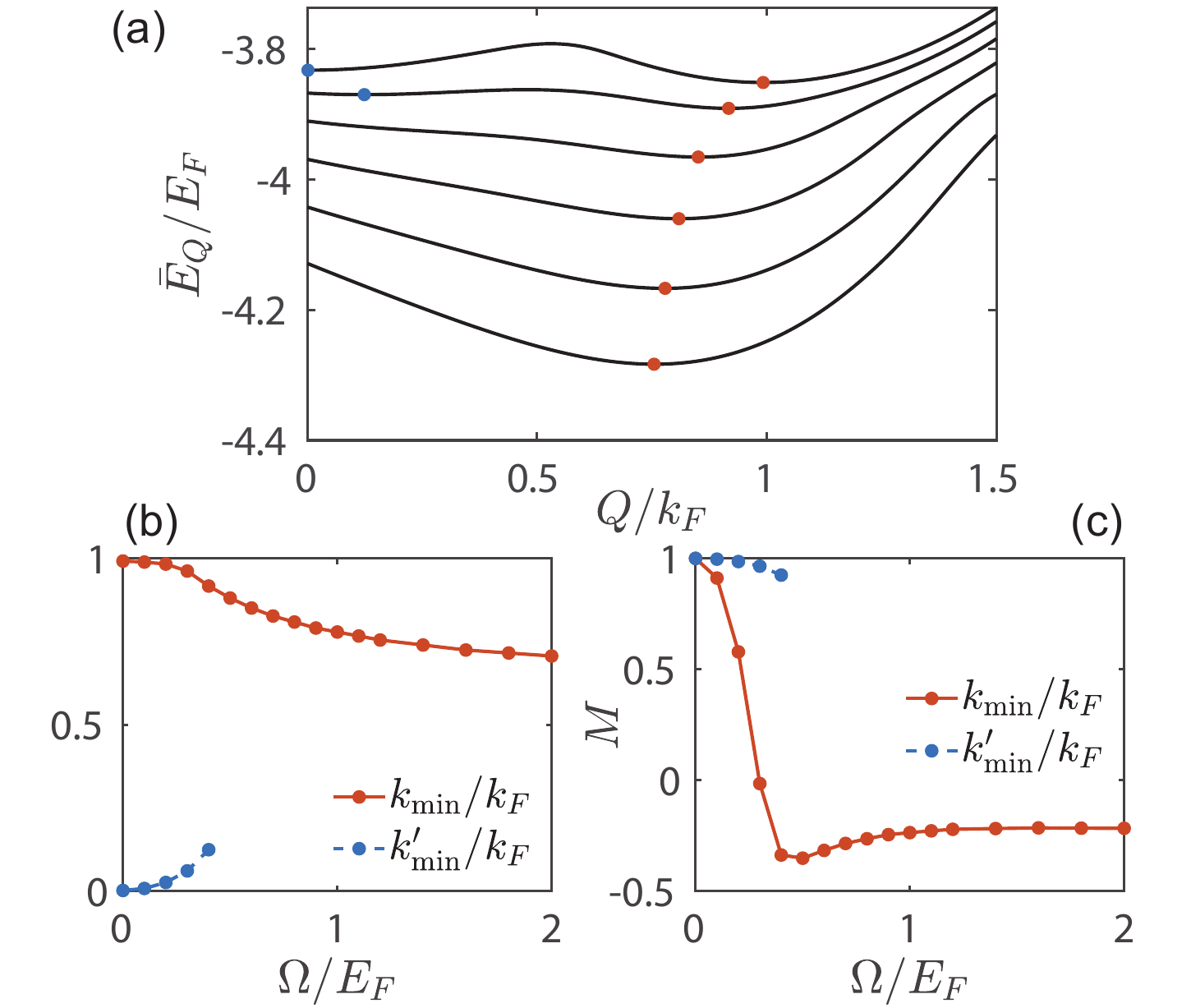}
\caption{(Color Online). Steady-state preparation of molecule state under spin-orbit coupling (SOC) with $\bar{k}=k_F$. Here $1/(k_Fa_s)=1.3$ and $\delta=E_{k_F}+0.05E_F$.  (a) Eigen-state dispersion $\bar{E}_{Q}$ of spin-orbit coupled polaron at different $\Omega/E_F=1.2, 1, 0.8, 0.6, 0.4, 0$ (from bottom to top). 
Red and blue points respectively mark the locations of  steady-state momentum $k_{\rm min}$ and the other disconnected minimum $k'_{\min}$.  (b) Evolutions of $k_{\rm min}$ and $k'_{\min}$ as changing $\Omega$. 
(c) Evolution of impurity magnetization $M$ as changing $\Omega$. The red and blue data respectively correspond to the minima at $k_{\min}$ and $k'_{\min}$. }  \label{fig_mol}
\end{figure}

Taking $1/(k_Fa_s)=1.3$ and $\delta=E_{k_F}+0.05E_F$,  we have numerically confirmed our scheme in Fig.\ref{fig_mol}. As shown by $\bar{E}_{Q}$ in Fig.\ref{fig_mol}(a), at large $\Omega$ the dispersion shows a single minimum at $k_{\rm min}$ (red point), which can be approached following the steady-state preparation under strong SOC. As reducing $\Omega$, the steady state adiabatically evolves and eventually stops at $k_{\min}=k_F$ when $\Omega\rightarrow 0$. The evolutions of $k_{\rm min}$ and $M$ for this steady-state are presented in Fig.\ref{fig_mol}(b,c). During this process, another minimum ($k'_{\min}$, blue point) emerges at an intermediate $\Omega$, which disconnects with $k_{\min}$ and continuously evolves to $0$ at $\Omega\rightarrow 0$, producing the polaron state.

\begin{figure}[t]
\includegraphics[width=8.5cm]{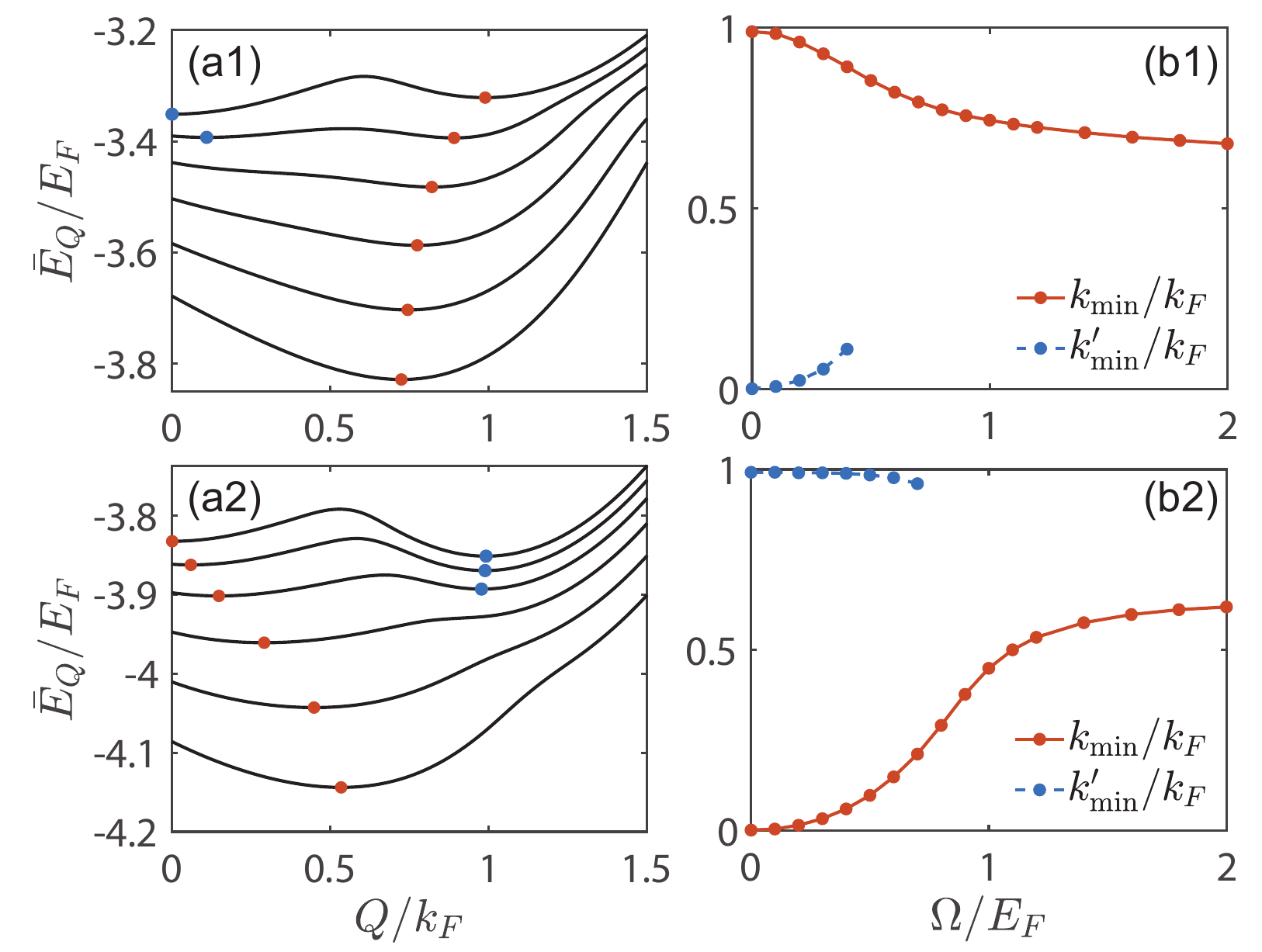}
\caption{(Color Online). (a1,b1) Same as Fig.\ref{fig_mol}(a,b) except for $1/(k_Fa_s)=1.2$ and $\delta=E_{k_F}+0.02E_F$, where the molecule is a metastable state with a higher energy than polaron. (a2,b2) Same as Fig.\ref{fig_mol}(a,b) except for a larger $\delta=E_{k_F}+0.4E_F$, where the steady-state evolution from large to zero $\Omega$ produces polaron state at $k_{\min}=0$.}  \label{fig_mol2}
\end{figure}

Our adiabatic scheme  equally applies to interaction regime where the molecule is metastable.  In Fig.\ref{fig_mol2}, we consider a weaker coupling $1/(k_Fa_s)=1.2$ when molecule  has a higher energy than polaron ($E_{k_F}>E_0$). Nevertheless, our results show that the adiabatic steady-state evolution of from large to zero  $\Omega$ can still produce molecule state, provided that the detuning $\delta$ is sufficiently close to $E_{k_F}$ (see Fig.\ref{fig_mol2}(a1,b1)). If this condition is not satisfied, such as for a larger $\delta$ far away from $E_{k_F}$ in Fig.\ref{fig_mol2}(a2,b2),  the evolution ultimately ends in the polaron state ($k_{\min}=0$) as $\Omega\rightarrow 0$. These distinct outcomes can be understood as follows: when $\delta\sim E_{k_F}$, the impurity exhibits sufficiently strong  correlation with states near $Q\sim k_F$ and finally the evolution  yields the molecule; when $\delta$ is far from $E_{k_F}$, this correlation becomes less important, while the impurity-polaron coupling at $Q\sim 0$ dominates, leading the adiabatic evolution to polaron. 

Given the robust underlying physics, we believe above phenomenon persists even when a more accurate ansatz with two particle-hole excitations is adopted. It has been shown that under such an improved ansatz, the polaron-molecule transition switches to $1/(k_Fa_s)_c\approx 0.9$\cite{Leyronas,Punk, Castin, Cui2}, with a coexistence window of $1/(k_Fa_s)\in (0.5,1.2)$\cite{Cui2}.  However, these quantitative improvements are not expected to  alter the feasibility of using SOC to probe molecule dispersions. In fact, the simple ansatz in (\ref{P_Q})  already yields $E_Q$ and $Z_Q$ near $Q\sim k_F$ with reasonable accuracy  in strong coupling regime ($1/(k_Fa_s)>1$), when compared to the improved ansatz\cite{Cui2}.  

Regarding the time scale for steady-state preparation, we note that a recent experiment reported a wide range of duration $t\sim 30 - 500\hbar/E_F$ to achieve the steady-state of Rabi-driven Fermi polaron\cite{Navon}. A crucial difference between Rabi coupling and SOC is that the latter can induce more atom losses due to two-photon Raman processes\cite{soc_review}. However, such atom loss can be well controlled in our setup because SOC is applied only to a small fraction of impurity atoms, not to the many-body background. Ideally, one can use post-selection to select out the single impurity subject to SOC.  
Alternatively, one can also reduce the SOC-induced losses by using  lanthanide\cite{Cui5, Lev}  or alkali-earth\cite{Ye} atoms.

Once the molecule state is prepared adiabatically, one can detect its dispersion near $Q\sim k_F$ as given by 
\begin{equation}
E_{Q}=E_{k_F}+\frac{Q_M^2}{2m_M^*}, 
\end{equation}
with $Q_M=Q-k_F$ the center-of-mass momentum\cite{Cui} and $m_M^*$ the effective mass. In realistic experiments with  a finite impurity density and at finite temperature $T$, $m_M^*$ can be extracted from the center-of-mass (or pair) momentum distribution of molecules, i.e., 
\begin{equation}
n(Q_M)\sim e^{-Q_M^2/(2m_M^*k_BT)}. 
\end{equation}
Such a pair momentum distribution has been measured previously in a 2D Fermi gas\cite{pair_distribution}. The successful detection of $m_M^*$ also requires that all molecules stay at thermal equilibrium around $Q\sim k_F$, rather than occupying other $Q$ regimes that do not follow the molecule dispersion. In practice, an  ideal situation for that is achieved at sufficiently low temperature $T<\Delta E$, where $\Delta E$ denotes the energy barrier between double minima in the dispersion curve. Near the polaron-molecule transition, $\Delta E\approx 0.05E_F$\cite{Cui2}, which imposes the requirement $T<0.05E_F$. We note that such low temperatures have already been realized in current Fermi polaron experiment\cite{Zwierlein2}.

Two remarks on molecule preparation are in order. First, the SOC momentum taken here ($\bar{k}=k_F$) is the most natural choice to build up correlation between the impurity (near $Q\sim 0$) and the molecule (near $Q\sim k_F$). If we take $\bar{k}\ll k_F$ instead, the impurity-molecule correlation is absent and the adiabatic process prefers to result in polaron rather than molecule. In this sense, the distinct $\bar{k}$ used to probe polaron and molecule clearly reflect their intrinsic momentum difference ($=k_F$)\cite{Cui,Cui2}, which is crucially important for understanding their first-order transition in single-impurity system. 
Secondly, the molecule prepared in the adiabatic process is practically very stable because of the directional symmetry breaking of its momentum (${\cp Q}$). Specifically, the direction of ${\cp Q}$ is uniquely pinned by that of $\bar{\cp k}$. In realistic system, such a unidirectional ${\cp Q}$ well protects molecules from tunneling to polarons due to the conservation of total momentum. This is very different from the conventional case where polarons and molecules can freely convert with each other in their coexistence regime\cite{Sagi, Cui}, basically because a pair of molecules at ${\cp Q}$ and $-{\cp Q}$ can  transit to two polarons at momenta near zero. In this way,   SOC enables a separate control of polaron and molecule by adjusting $\bar{k}$, thereby facilitating the exploration of their individual dispersion in different momentum regions. 

\bigskip

\section{Summary and outlook} \label{summary}

In summary, we have shown that SOC can be used as an efficient tool to probe Fermi polaron and molecule dispersions. Specifically, the dispersions can be either detected through the linear response of impurity spin at small SOC momentum $\bar{k}$, or via an adiabatic steady-state evolution at large $\bar{k}=k_F$. The distinct $\bar{k}$  reveal a fundamental momentum difference between polaron and molecule, therefore offering a clear picture for their first-order transition and coexistence. Moreover, the adiabatic scheme is able to prepare stable molecules with directional symmetry breaking, which facilitates an individual manipulation of this state even in polaron-molecule coexistence regime. 

Our theory  can also be applied to unequal-mass case, provided that the mass ratio $m_f/m$ is below the critical value for trimer formation\cite{KM,Pricoupenko}, such that  two-body correlation still dominates to support the polaron-molecule transition\cite{Cui3, Cui4}. A typical example of unequal-mass system is the Li-K mixture, in which polaron physics has been extensively studied experimentally\cite{Grimm, Grimm2016, Grimm2021, Grimm2024}, and our scheme can be directly applied. Finally, our detection scheme  is expected to be work equally well  in 2D, where polaron-molecule transition persists\cite{Parish, Parish2, MC_2d_1, MC_2d_2} and the impurity dispersion exhibits a similar double-minimum structure\cite{Cui2}.


\bigskip

Data that support the findings of this article are openly available~\cite{data}.

\bigskip

{\it Acknowledgements.} This work is supported by National Natural Science Foundation of China (12525412, 92476104, 12134015) and Quantum Science and Technology-National Science and Technology Major Project (2024ZD0300600).

\end{document}